# Beam-Transport Systems for Particle Therapy

*J.M. Schippers*
Paul Scherrer Institut, Villigen, Switzerland

**Abstract**
The beam transport system between accelerator and patient treatment location in a particle therapy facility is described. After some general layout aspects the major beam handling tasks of this system are discussed. These are energy selection, an optimal transport of the particle beam to the beam delivery device and the gantry, a device that is able to rotate a beam delivery system around the patient, so that the tumour can be irradiated from almost any direction. Also the method of pencil beam scanning is described and how this is implemented within a gantry. Using this method the particle dose is spread over the tumour volume to the prescribed dose distribution.

## 1 Introduction

The main purpose of the beam-transport system is to aim the proton beam, with the correct diameter and intensity, at the tumour in the patient and to apply the correct dose distribution. The beam transport from the accelerator to the tumour in the patient consists of the following major sections (see Fig. 1):

– energy setting and energy selection (only for cyclotrons);

– transport system to the treatment room(s), including beam-emittance matching;

– per treatment room—a gantry or a fixed beam line aiming the beam from the correct direction;

– beam-delivery system in the treatment room, by which the dose distribution is actually being applied. These devices are combined in the so called 'nozzle' at the exit of the fixed beam line or of the gantry.

In the last few years, there have been many developments by commercial companies, to supply a facility with only one treatment room. Most of the contents of this chapter also apply to these facilities. The only exception is the version in which the beam from a cyclotron is sent to the patient, directly. In that case, the cyclotron is immediately followed by the nozzle components.

The details of the components in the nozzles are discussed in other chapters in these proceedings.

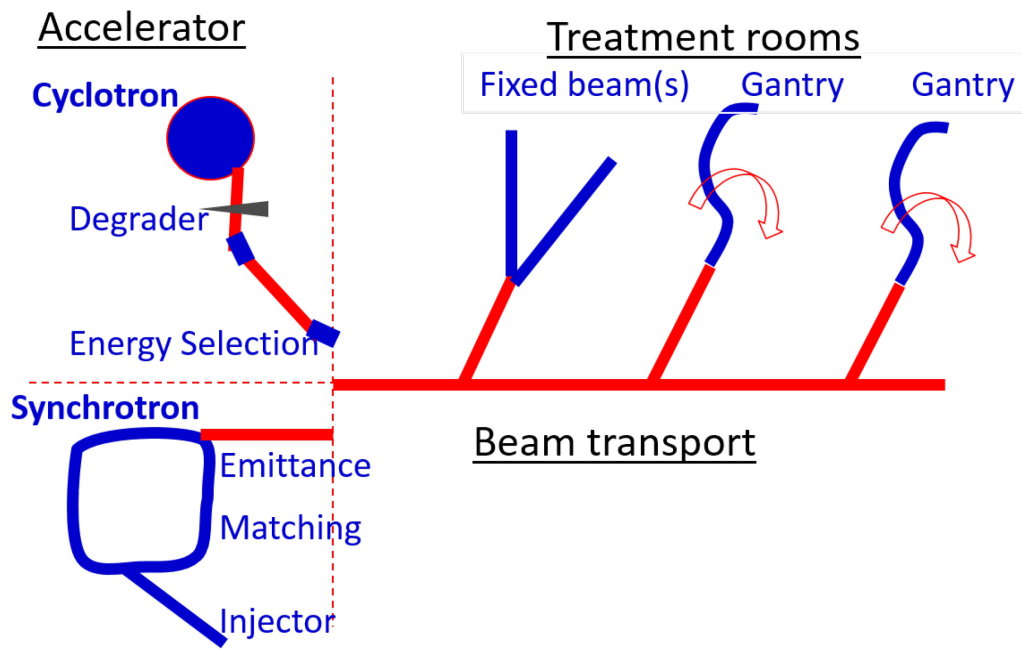

**Fig. 1:** A schematic overview of the accelerator and beam-transport system in a particle-therapy facility. Both currently used accelerators are indicated. Drawing is not to scale.

There are two major techniques for applying the dose to the patient. After aiming the beam in the desired direction by rotating the gantry (beam-transport system mounted on a rotating structure) to the correct angle, the beam must be spread in the lateral direction since the tumour diameter is larger than the beam cross section. Therefore, one must the match beam shape to the cross section of the tumour, as seen from the incoming beam direction. This is done either by the scattering technique or the scanning technique. In the scattering method, the beam cross section is increased by sending the beam through a system by which the beam diameter is increased by scattering in a foil system (a 'passive' technique) to match to the maximum lateral tumour dimension. In the scanning method, a 'pencil beam' with a diameter of approximately 1 cm is 'actively' scanned in the transverse plane over the tumour cross section. This motion is done in steps, and the applied 'spot' dose is varied per step ('spot scanning'). In a method currently in development, scanning is performed by a continuous shift of the beam along lines in the tumour. During this sweep, the beam intensity is varied to deliver the correct dose along the line ('continuous scanning'). Until now, the scattering technique has most commonly been used. However, for several years, the scanning technique has been regarded as the optimal technique (i.e. the technique that applies the best possible dose distribution) currently feasible in practice and almost all new facilities are designed to employ this technique. Therefore, in this chapter, the focus will be on the application of scanning techniques.

In this technique, the beam-transport system is controlling the following parameters:

– beam position by scan magnets;
– beam size by focusing;
– beam energy (in a synchrotron or in a degrader following a cyclotron);
– beam intensity by controlling the intensity from the accelerator, or by controlling the beam transmission in the beam-transport system.

## 2 Beam transport

### 2.1 Setting of the beam energy

As mentioned in the Introduction, the beam transport between the accelerator and treatment rooms has different functions. In most facilities, these functions are performed in different sections.

In the case of a cyclotron, first, the beam energy is set to the value needed at the treatment site. A choice is made concerning how quickly this system should set the energy. If the beam energy is only set to the value that corresponds to the maximum needed range in the patient (= the deepest part of the tumour as seen from the beam direction), this value does not need to change very quickly (fraction of a minute), since it can be done, for example, during gantry rotation. In that case, the range modulation over the thickness of the tumour is done in the nozzle by using range-shifter plates or a range-modulation wheel, in case scattering is used as the dose-application technique, see Fig. 2.

If a synchrotron is used as the accelerator, each spill is extracted with a specific energy. The waiting time until the next spill is of the order of several seconds, so this can be used, conveniently, to set the maximum energy and perform the modulation in the nozzle. Recent developments in synchrotrons also enable an energy change during a spill [1]. Although not yet implemented, this is a very promising development.

In the case of a cyclotron, the energy modulation can be done at the start of the beam-transport system. This is quite challenging. Apart from the degrader, all following magnets in the beam transport and gantry must be able to make rather fast changes. A high-speed energy change will quickly vary the Bragg-peak position, in depth, and this will limit the time of the total dose application. During modulation of the proton energy, a typical energy step corresponds to a range step of approximately 5 mm in water. This corresponding step of approximately 1% in momentum of the proton beam must be made in a fraction of a second. In that case, the mechanical design of the degrader must be such that the required speed can be reached, but also, the following magnets in the beam line and gantry must be able to change their field accordingly. This requires power supplies that can change the magnet current quickly and use a voltage overshoot to compensate the induction of the magnet.

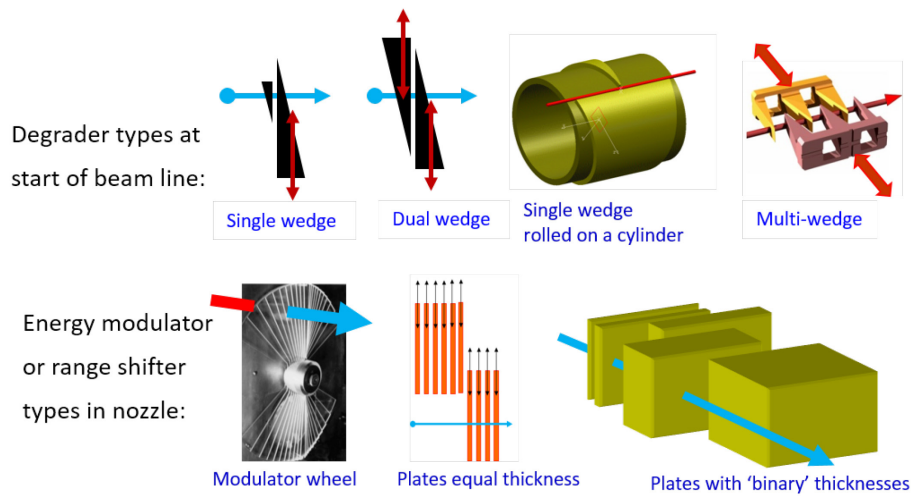

**Fig. 2:** Top: Several possible degrader systems consisting of wedges that vary the thickness of the material to be crossed by the beam. Bottom: a wheel that modulates the beam energy when it rotates, and two systems of plates that can quickly be put into/out of the beam line for fast energy (range) adjustments. The plates could also be installed as degraders at the beginning of the beam line.

The degrader usually consists of a system of Lucite or graphite wedges, and different mechanical possibilities are shown in Fig. 2. The amount that the wedge(s) has been shifted into the beam line determines the amount of material that is traversed by the beam, and thus, the energy of the ongoing beam. Different mechanical variations of the wedge are in use. The wedge can be rolled along the surface of a cylinder, or a system of multiple wedges can be used. In all cases, the position of the edge(s) must be set quickly and accurately, and at least two wedges are needed to obtain a uniform thickness over the beam diameter.

For a given cyclotron-beam energy, the minimum energy of the particles leaving the degrader depends on the total thickness of the traversed material. Usually, one degrades until a beam energy of approximately 70 MeV is reached. At lower energies, the beam transport could be distorted (become un-sharp) too much by multiple scattering in, for example, the vacuum window at the exit of the beam line and dosimetry devices just before the patient. The maximum energy of the continuous energy variation range is, usually, approximately 20 MeV below the energy from the cyclotron. This is due to the minimum overlap of the wedges, which needs to be the same thickness over the beam cross section.

There will be a spread in beam energy of the particles that leave the degrader. This energy straggling is caused by the statistical variation of the particle tracks in the degrader. This spread increases with the energy loss in the degrader and can be larger than the energy acceptance of the beam-transport system. Therefore, the degrader is followed by an energy selection system (ESS), consisting of a dipole followed by a slit system. In the ESS, a maximum relative (%) beam-momentum spread is selected.

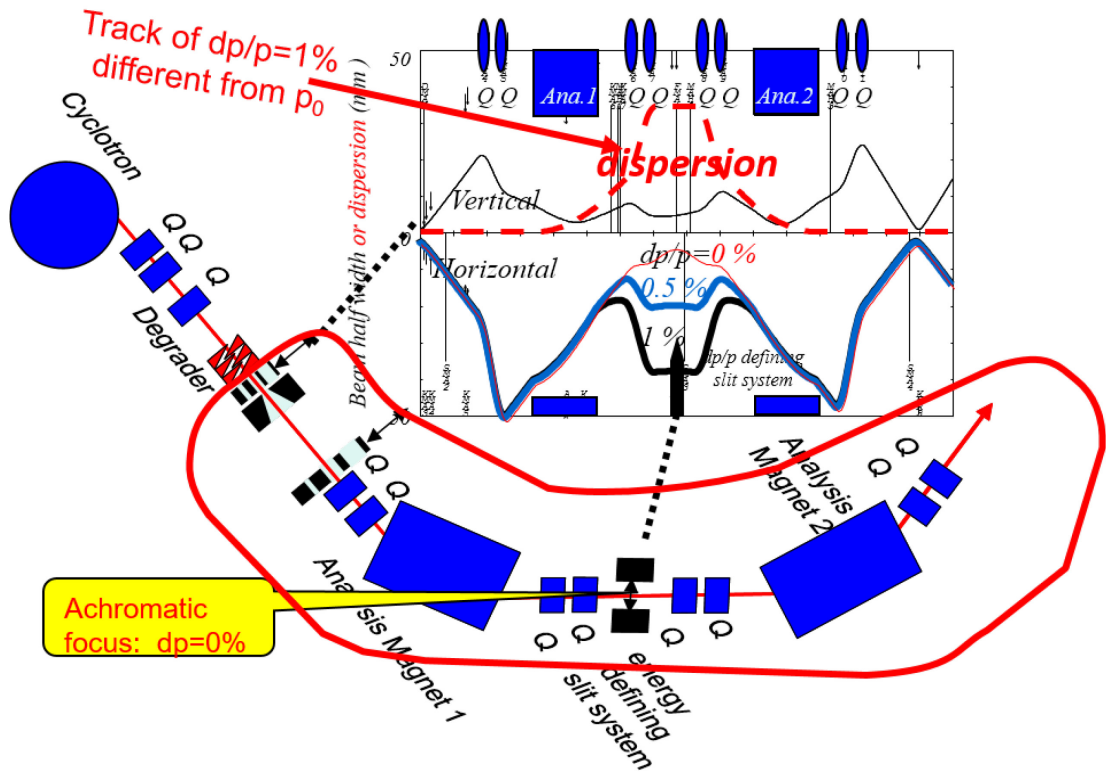

**Fig. 3:** The beam optics in the ESS. In the top graph, the half beam size is plotted as a function of the position along the beam line. Above the *x*-axis, the vertical dimension is plotted as a solid line and the dispersion as a dashed line. Below the *x*-axis, the horizontal beam size is plotted for 0%, 0.5% and 1% momentum spread. The black arrows indicate the momentum acceptance slit aperture.

As shown in Fig. 3, the beam optics in the ESS are usually made such that there is a large dispersion at the slit position ('dispersion' is the track of a particle with 1% momentum deviation). But, to get a well resolved energy-position dependence at one's position, one also needs a monochromatic focus at one's position. This is the case when there would be a small beam width if the beam were without energy spread (the 0% dp/p line in Fig. 3). This combination of dispersion and monochromatic focusing results in a beam profile in which there is a unique relation between horizontal position and energy. Then, the total width is mostly determined by the energy spread in the beam, as shown in Fig. 4.

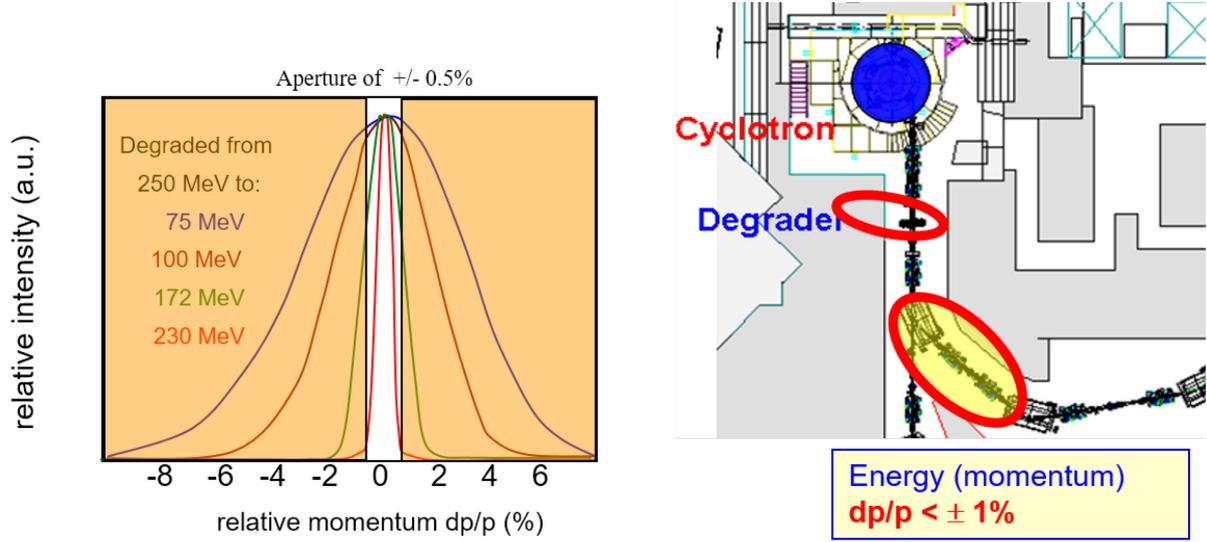

**Fig. 4:** Left: the momentum spread, which almost equals the beam spread in the horizontal plane, at the momentum-selecting slit. The aperture used, of e.g. 0.5%, has been indicated. Right: the location of the degrader and the ESS in the PSI proton-therapy facility.

Due to the momentum selection at the slit, the beam energy behind the ESS is mainly determined by the magnetic field of the bending magnet. The wrong magnetic field will send the incorrect energy through the slit system, yielding the wrong range within the patient. The relation is

$$\frac{dR}{R} = 3.2 \frac{dB}{B}, \qquad (1)$$

so that an error of 1% in the magnetic field $B$ would lead to a range error of 6.4 mm, at a (water equivalent) depth of $R = 20$ cm in the patient. Therefore, the field setting must be reproducibly correct within $10^{-4}$. The absolute degrader setting is less critical ($<10^{-2}$), due to the width of the energy straggling. A small setting error will only lead to a shift of the profile at the slit (see Fig. 4), yielding a slightly reduced transmission through the ESS.

Using the aperture in the slit system, one thus selects a certain momentum spread, typically, approximately, 0.5–1%. This is also the momentum spread at the patient. One should realize that this momentum spread is smaller than in the case of energy degrading (to modulate the range) being performed in the nozzle, since, in that case, no energy selection is done after crossing the range-shifter plates or the modulation wheel. The smaller momentum spread after an ESS will give sharper Bragg peaks in the patient. As shown in Fig. 5, especially for low energies, this has consequences for the sharpness of the Bragg peak, which must be taken into account in treatment planning.

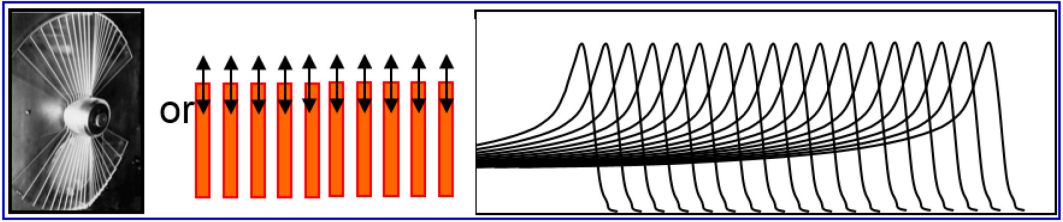

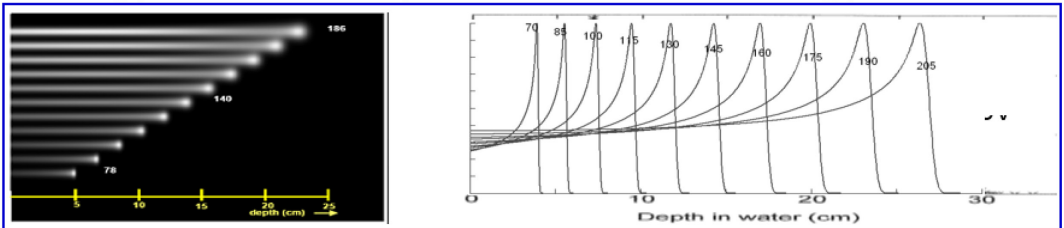

**Fig. 5:** The shape (sharpness) of the Bragg peak in the patient for different energy-reducing methods. Top: methods used in the nozzle, where no energy selection is done. Bottom: a degrader, followed by an ESS, will give sharp Bragg peaks at low energies.

The degrader not only decreases the beam energy; it also increases the emittance due to multiple scattering. In order to minimize this effect, but while keeping sufficient stopping power, a material of low atomic number $Z$ is used. However, the emittance increase is considerable, and the degrader must be followed by a set of apertures that limit the emittance to the acceptance of the following beam line. Since the emittance also increases with decreasing energy out of the degrader, the fraction of the beam that is accepted in the beam transport, behind these collimators and the ESS, is also decreasing, the lower the energy. Figure 6 [2], shows the order of magnitude of this transmission loss at the 250 MeV beam from the cyclotron at PSI.

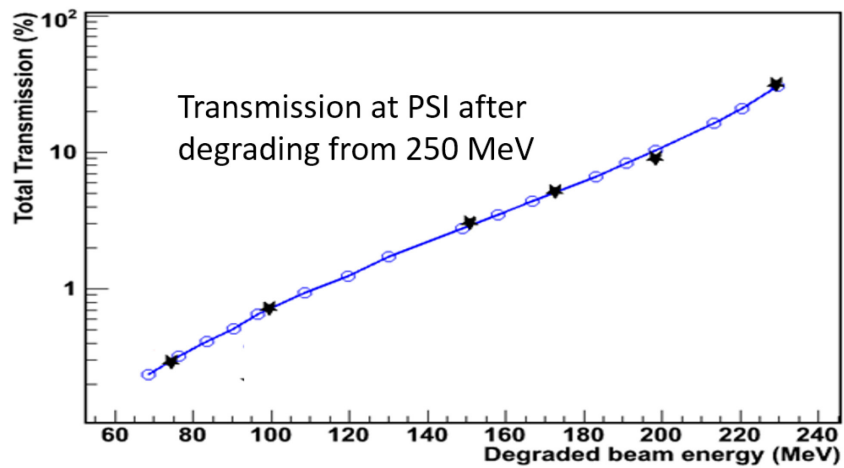

**Fig. 6:** The transmission of the proton beam in the ESS and beam-transport system, as a function of the energy behind the degrader, for a beam of 250 MeV from the cyclotron [2].

The beam intensity at the patient would be strongly dependent on beam energy, due to this energy-dependent transmission. This would complicate the treatment and could have consequences for safety, due a limitation of reaction times. Therefore, an intensity compensation is necessary to obtain a beam intensity that is not so dependent on energy. This can be done by adjusting the intensity in the cyclotron, but one could also design a beam-line setting with energy-dependent controlled beam losses at dedicated collimators in the beam-transport system.

Of course, all beam losses will create activation. Therefore, it is important to concentrate the beam losses at well-known locations. These can be shielded if necessary, and also, one can select materials that will not contain long-lived radioactive isotopes. In a cyclotron facility, the degrader, acceptance defining collimators, and the ESS are usually the only such locations in the beam-transport system. The losses due to intensity compensation are relatively low.

## 2.2 Optics of the beam transport

In a synchrotron, the beam energy is set in the ring by acceleration until the desired energy has been reached. A change to other energies thus requires a new spill, which takes a few seconds. Therefore, in synchrotron facilities, energy modulation is usually done in the nozzle. Therefore, scattering is still the method most commonly used in synchrotron facilities. However, as discussed in the chapter on synchrotrons (these proceedings), there have been very interesting recent developments, aimed at beam-energy reduction during a spill.

Another difference with cyclotrons is the extracted emittance. The emittance of the beam from a cyclotron is slightly asymmetric (horizontally versus vertically), but this asymmetry is overruled by the emittance increase in the degrader, which is symmetric. The horizontal emittance of a synchrotron beam can be a factor of ten smaller than in the vertical direction. This would give a gantry-angle dependence in the dose-application process. Therefore, emittance matching is necessary if a gantry is used. This can be done by a system of rotating quadrupoles that rotate the emittance with the gantry (see chapter on this topic), or by passing the beam through a scatter foil to make a symmetric emittance.

The further optics of the beam transport should have a layout that is as stable as possible, thus providing a high reproducibility of the beam-transport characteristics. Since the emittance of a degraded beam from a cyclotron is much larger than that of a synchrotron, the aperture of most magnets is usually larger in the case of a cyclotron. However, apart from the degrader with an ESS and the emittance matching, the beam optics are rather similar for both types of machines, as already indicated in Fig. 1. A beam transport with several intermediate images is very stable and reproducible and easy to verify by profile monitors at these locations. Collimators, at one or more of these locations, also convert beam alignment errors into an intensity reduction, which is usually less of a problem.

One should be able to rely on the reproducibility of the beam-line setting. Therefore, the temperature of the vaults and the cooling water should not fluctuate. Also, one should always use the same procedures for, e.g., field ramping and beam-line settings for other energies. During or in-between patient treatments there is no possibility to tune (fiddle with) the beam or to try a new setting.

Beam-diagnostic tools, such as profile monitors, should only be used when developing new beam-line settings or in quality assurance (QA) programs. They should be removed from the beam trajectory and this position should be continuously verified. A non-interceptive beam-diagnostic tool is an advantage and can be used during treatments.

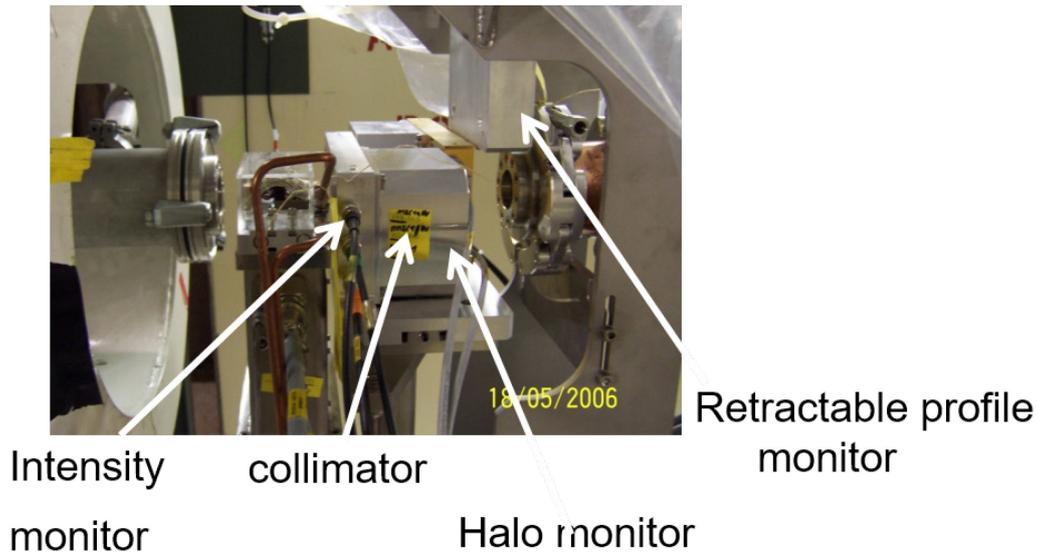

**Fig. 7:** The gantry coupling point at PSI. The beam direction is from right to left. In an air gap of 20 cm, a collimator and beam-diagnostic tools have been mounted. The non-interceptive intensity and halo monitors continuously provide the control system information, including during patient treatment.

For example, at PSI such beam-diagnostic tools have been mounted at the coupling points in all treatment rooms and are always active [3]. Here, the vacuum of the beam line is separated from the vacuum of the gantry by an air gap of approximately 20 cm, as shown in Fig. 7. A retractable profile monitor must be retracted during treatments. However, the ionization of the air in the gap behind this monitor is used to measure the beam intensity as well as the presence and orientation of the beam halo. There are no foils in the beam. Electric fields are guiding the charge to foils next to the beam path.

## 3    Gantry

In ion-therapy facilities, normally, use is made of different fixed beams, using beam directions in the horizontal direction, in the vertical direction, and/or from an inclined direction. In combination with a slightly flexible patient-positioning system, beam directions around the fixed directions can be used. But, in order to direct the particle beam from all different directions to the tumour, other methods are in use. A gantry offers most flexibility [4].

When using a gantry, the beam-transport system ends at a coupling point to the gantry. Downstream of this point, the beam-transport system is mounted on a mechanical structure, which can rotate 360 degrees, or a bit more than 180 degrees, around the patient. At the coupling point, the beam should be symmetric, both in shape and in divergence, so that there is no dose-application dependence on the gantry angle. In a gantry, the beam must be bent several times to get the correct beam, which is perpendicular to the incoming beam direction. The magnetic rigidity of the protons implies the use of strong (>1.5 T) and large magnets. A distance of at least 2 m is needed in order to provide the necessary distance for a scattering system and the length to spread the scattered beam. For pencil-beam scanning, a similar amount of space is needed for the scanning magnets and the length to get sufficient lateral displacement of the scanned beam. Also, space is needed for dose monitors, range-shifter and scanning magnets, or energy modulator and collimation systems between the exit of the last magnet and the patient. Therefore, the typical diameter of these 100 ton gantries is 10–11 m, both for scattered beams and for scanned beams (see, for example, Fig. 8).

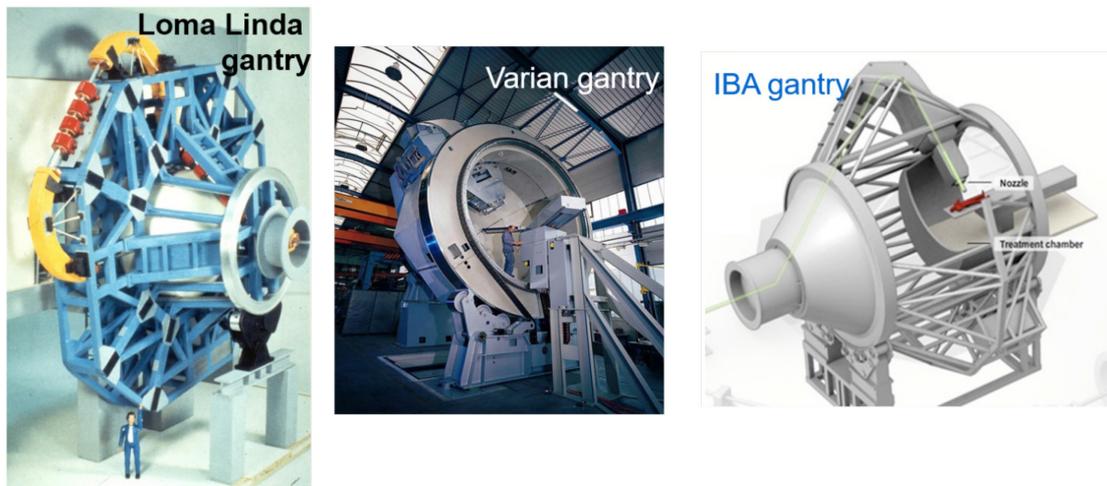

**Fig. 8:** Left—the 'corkscrew' gantry at Loma Linda; middle—the Varian gantry designed for scanning beams during its construction phase; and right—the IBA gantry.

Due to the almost three times larger magnetic rigidity of carbon ions (approximately 6.8 Tm), the radius of curvature of particle trajectories is more than 3 m with conventional magnets. The huge difficulties and costs associated with a carbon gantry have resulted in only one existing gantry for carbon ions (13 m diameter, 570 tons), which has been installed at the HIT facility in Heidelberg, see Fig. 9 [5]. However, a gantry with superconducting magnets has been designed and built for the extension of the Japanese ion-therapy facility HIMAC [6].

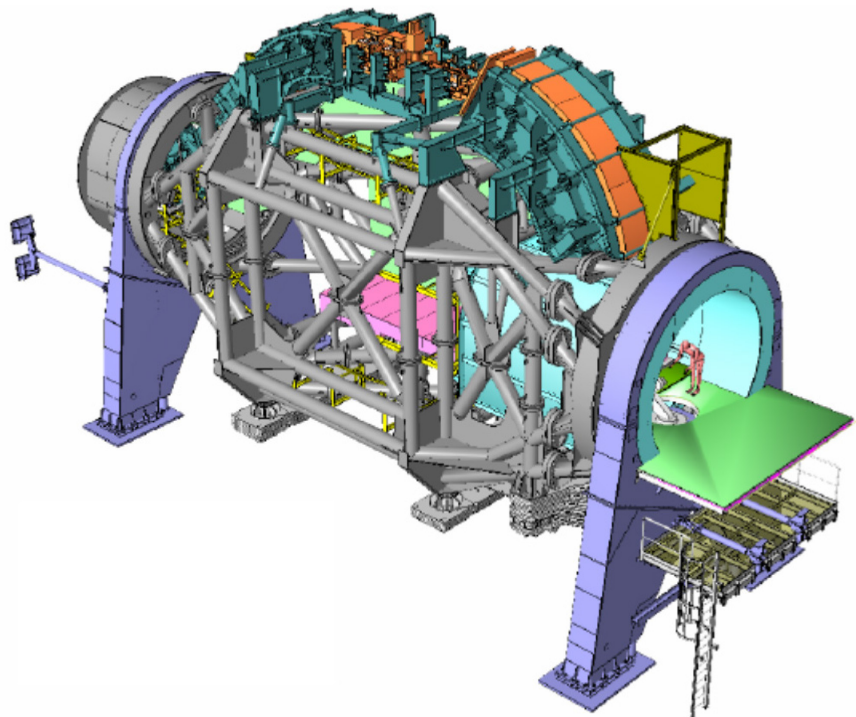

**Fig. 9:** The gantry at HIT in Heidelberg for carbon therapy

The first three gantries in particle therapy originated from a design by the Harvard group [7] and were installed at the facility in Loma Linda. The beam-transport-system layout is referred to as a *corkscrew* and consists of a double-achromatic-bending system, resulting in a rather short gantry design, as shown in Fig. 8. The nozzle has been designed for a scattered beam. But, with the increasing demand for pencil-beam scanning, dedicated nozzle designs have been made to allow (also or only) scanning. In the case of scanning, the scanning magnets have usually been mounted at the exit of the last bending magnet ('downstream scanning'). Their location acts as the virtual source of the pencil beams.

At PSI, a compact proton gantry ('Gantry1') has been in use since 1996 [8], which is optimized for pencil-beam scanning. In this gantry, a scanning magnet has been mounted before ('upstream scanning') the last bending magnet (90 degrees), which bends the beam towards the isocentre. The beam optics has been designed such that a deflection of the pencil beam by the scanning magnet causes a parallel shift of the beam at the isocentre. In this way, the virtual source of the pencil beams is located at infinity and an orthogonal pencil-beam arrangement is obtained. The other two orthogonal displacements are performed by inserting range-shifter plates in the nozzle to shift the Bragg peak in depth and by shifting the table in the direction orthogonal to the magnetic scanning direction. The maximum speed of the dose application is limited by the necessary slow motion of the patient table.

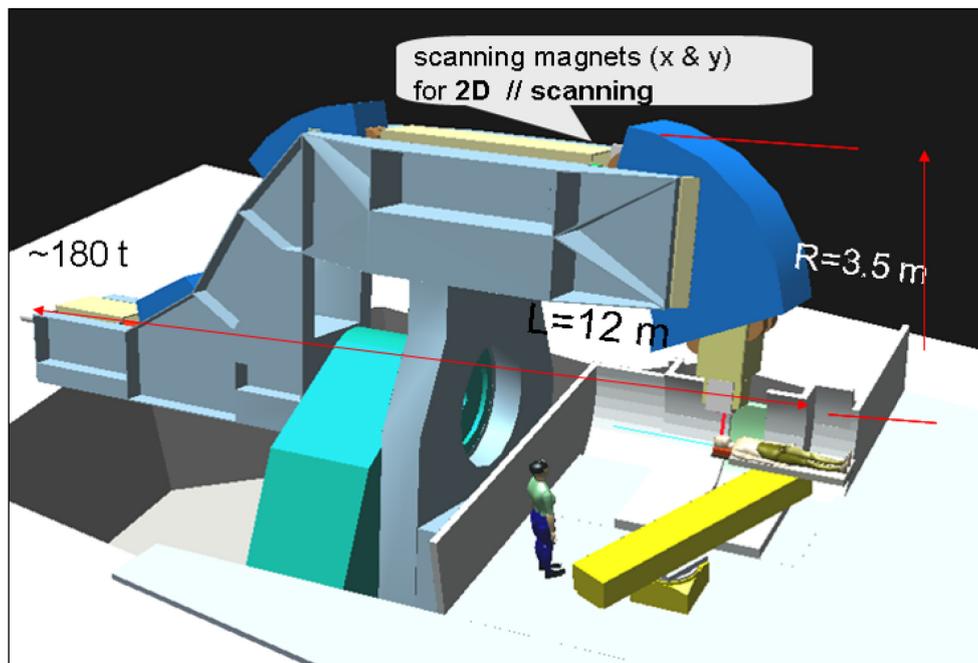

**Fig. 10:** Design of Gantry 2 at PSI. This gantry is designed for parallel beam scanning in two dimensions and fast energy scanning.

In order to have fast (i.e. magnetic) scanning in two directions, 'Gantry2' has been designed and built at PSI, see Fig. 10 [9]. This gantry allows for double magnetic scanning in a field of $12 \times 20$ cm$^2$, with parallel beam displacements in both directions. To allow scanning over 12 cm in the direction orthogonal to the bending plane of the last (also 90 degrees) bending magnet, a relatively large gap of this bending magnet is needed. Since this gap limits the field size in this direction, a table shift of <12 cm is necessary once or twice at a certain gantry angle. The magnets of Gantry 2 are laminated, so that their magnetic fields can change with the energy of the incoming beam. With this combination of laminated magnets in the gantry and in the preceding beam line following a degrader, a range change of ~5 mm in water is achieved in ~80 ms.

A gantry design employing a beam-focusing concept normally used in FFAG (Fixed Field Alternating Gradient) accelerators, allows a very large momentum acceptance of $\pm 15\%$ [10]. The design of the magnet system shows very-tightly packed focusing and defocusing magnets, with gradients up to 70 T/m to be achieved with superconducting magnets. This has not been built yet, however, but it has triggered several groups to work, also, on various other designs of gantries with superconducting magnets [11].

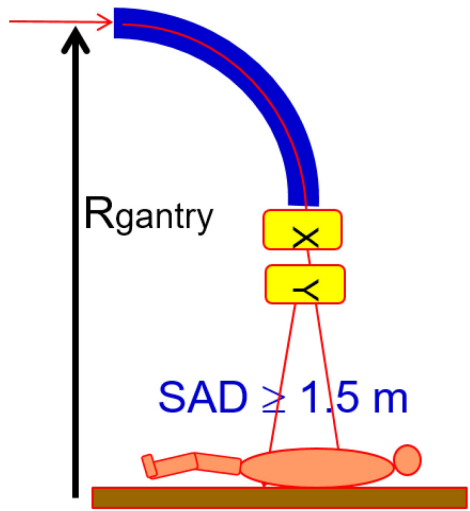 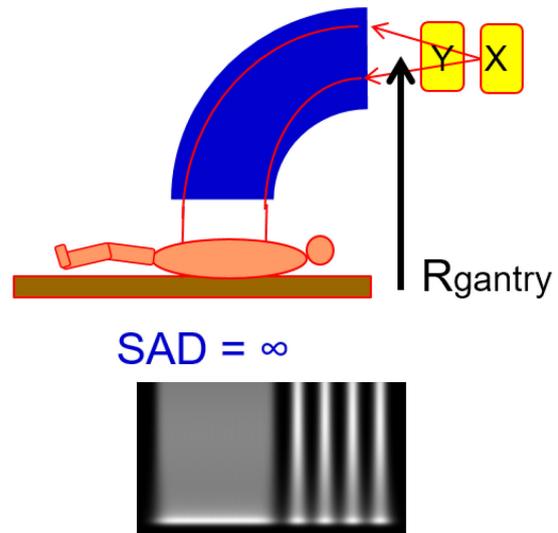

**Fig 11:** Schematic overview of the two most used scanning magnet configurations: before ('upstream') or behind ('downstream') the last bending magnet. SAD= source–axis distance.

So, when implementing the scanning technique, one can make a choice between upstream scanning, downstream scanning (see Fig. 11), or a combination of the two [12]. The advantage of downstream scanning is the simpler layout of the large final bending magnet. However, due to the needed space behind the last bending magnet, the gantry radius will not be small.

In the case of upstream scanning, a large aperture and more complicated optics are needed to obtain appropriate 2D pencil-beam motions in combination with sufficient focusing of the pencil beams. The gantry radius can be made rather small, since only approximately 1 m of space is needed for dosimetry equipment, etc. Upstream scanning also offers the possibility to get a parallel displacement of the pencil beam, which has several advantages. This is done via a so-called point-to-parallel imaging, by means of the inclined angles of the entrance and exit of the last bending magnet.

## 4   Conclusions

The beam transport between accelerator and gantry has different purposes. The most important is to aim the proton beam with the correct diameter and intensity at the tumour in the patient and to apply the correct dose distribution. This task is achieved by performing several actions, usually performed in clearly distinguishable beam-line sections. For example, these could be a section for setting the energy (in the case of a cyclotron), a section to perform the transport of the beam to the treatment room(s), and a final section that aims the beam at the tumour from the correct direction.

An accelerator facility for particle therapy implements a variety of technical measures to ensure an accurate and reproducible dose delivery to patients. In comparison to an accelerator for physics research, a medical irradiation facility differs mostly in relation to reliability and simplicity of the equipment and operational procedures. Certain measurements need to be performed frequently, so that absolute dose values are delivered to the patients with an accuracy of a few percent. The necessary safety measures related to the irradiation treatment impose a stringent discipline, and procedure-following regulations, on the operation and implementation of changes or upgrades.

## Acknowledgement

I would like to thank the numerous colleagues at other particle-therapy institutes and at several companies for the countless discussions at conferences and workshops and the detailed information given at site visits.

## References

[1] Y. Iwata *et al.*, Multiple-energy operation with quasi-dc extension of flattops at HIMAC, MOPEA008, Proc. IPAC'10, Kyoto, Japan, 2010, p. 79.
   http://epaper.kek.jp/IPAC10/index.htm

[2] M.J. van Goethem *et al.*, *Phys. Med. Biol.* **54** (2009) 5831.
   http://dx.doi.org/10.1088/0031-9155/54/19/011

[3] R. Dölling et al., Beam diagnostics for the proton therapy facility proscan, Proc. AccApp'07, 2007, p. 152.

[4] J.B. Flanz, Proc. PAC95, Dallas, TX, USA, May 1–5, 1995, p. 2004.
   http://accelconf.web.cern.ch/AccelConf/p95/

[5] Th. Haberer *et al.*, *Radiother. Oncol.* **73** S186 (2004).
   http://dx.doi.org/10.1016/S0167-8140(04)80046-X

[6] K. Noda *et al.*, New heavy-ion cancer treatment facility at HIMAC, TUPP125. Proc. EPAC08, 2008, p. 1818.

[7] A.M. Koehler, Proc. 5th PTCOG Meeting: Int. Workshop on Biomedical Accelerators, Lawrence Berkeley Laboratory, Berkeley, CA, 1987, p. 147.

[8] E. Pedroni *et al.*, *Med. Phys.* **22(1)** (1995) 37. http://dx.doi.org/10.1118/1.597522

[9] E. Pedroni *et al.*, *Z. Med. Phys.* **14(1)** (2004) 25. http://dx.doi.org/10.1078/0939-3889-00194

[10] D. Trbojevic *et al.*, Proc. PAC07, Albuquerque, New Mexico, USA, June 25-29, 2007, p. 3199.
   http://epaper.kek.jp/p07/INDEX.HTML

[11] W. Wan *et al.*, Alternating gradient canted cosine theta superconducting magnets for future compact proton gantries, Phys. Rev. ST—Acc. and Beams 18, 103501 (2015)

[12] H. Vrenken *et al., Nucl. Instr. Meth.* **A 426(2–3)** (1999) 618. http://dx.doi.org/10.1016/S0168-9002(99)00039-X